\documentclass{iopart}
\usepackage{iopams,setstack}
\newcommand{\btimes}{\mbox{\boldmath$\,\times\,$}}
\begin{document}

\title{Comment on `Electromagnetic force on a moving dipole'}
\author{V Hnizdo}

\address{National Institute for Occupational Safety and Health,\\
Morgantown, West Virginia 26505, USA}
\ead{vhnizdo@cdc.gov}
\begin{abstract}
Using the Lagrangian formalism, the force on  a moving dipole derived by Kholmetskii, Missevitch and Yarman (2011 {\it Eur.\ J.\ Phys.} {\bf 32} 873--81) is found to be missing  some important terms.
\end{abstract}

\noindent
Recently, Kholmetskii, Missevitch and Yarman  \cite{KMY}  have inspected the analysis of Vekstein  \cite{Vek} of the force on a small system  of zero net charge but with electric and magnetic dipole moments $\bi d$ and $\bi m$, respectively, moving with velocity $\bi v$ in an electromagnetic field ${\bi E},{\bi B}$. They concluded that
the last term in Vekstein's expression for the force,
     \begin{equation}
{\bi F}_{\rm V}=({\bi d}\bdot\bnabla){\bi E}
+\bnabla({\bi m}\bdot{\bi B})+\frac{1}{c}\, {\bi d}\btimes
({\bi v}\bdot \bnabla){\bi B},
\label{F_V}
\end{equation}
is erroneous and derived for the force the following expression:
\begin{equation}
{\bi F}_{\rm KMY}
=\bnabla({\bi d}\bdot{\bi E})
+\bnabla({\bi m}\bdot{\bi B})+\frac{1}{c}
\frac{\partial}{\partial t}\,({\bi d}\btimes{\bi B}).
\label{F_KMY}
\end{equation}
In both (\ref{F_V}) and (\ref{F_KMY}), the dipole moments $\bi d$ and $\bi m$
may arise from the rest-frame dipole moments  ${\bi m}_0$ and
${\bi d}_0$ according to  transformations
that read to first order in $v/c$ as \cite{PP}\footnote{The magnetic moment ${\bi m}=-{\bi v}\btimes{\bi d}_0/c$ of a moving rest-frame electric dipole ${\bi d}_0$ is not  the standard magnetic dipole moment $(1/2c)\int{\rm d}^3r\, {\bi r}\btimes {\bi J}({\bi r})$  of a divergenceless current distribution ${\bi J}({\bi r})$.
It is to be regarded as a quantity the correct use of which will yield the force on the moving dipole exerted by an external magnetic field.}
\begin{equation}
{\bi d} = {\bi d}_0
+ \frac{1}{c}\,{\bi v}\btimes{\bi m}_0, \quad\quad
{\bi m} = {\bi m}_0
- \frac{1}{c}\,{\bi v}\btimes{\bi d}_0.
\label{d,m}
\end{equation}

In this comment, we employ the Lagrangian formalism to obtain the force on a zero-charge particle with rest-frame electric and magnetic dipole moments, moving in a static electromagnetic field. The resulting force, correct to first order in $v/c$, is
\begin{eqnarray}
\fl {\bi F}&=\bnabla({\bi d}\bdot{\bi E})
+\bnabla({\bi m}\bdot{\bi B})
-\frac{1}{c}\,{\bi m}_0\btimes ({\bi v}\bdot \bnabla){\bi E}
+\frac{1}{c}\, {\bi d}_0\btimes
({\bi v}\bdot \bnabla){\bi B}\nonumber\\
\fl &\quad
-\frac{1}{c}\,\dot{\bi m}_0\btimes {\bi E}
+\frac{1}{c}\,\dot{\bi d}_0\btimes {\bi B}.
\label{F}
\end{eqnarray}
Here, the term
$(1/c)\, {\bi d}_0\btimes({\bi v}\bdot \bnabla){\bi B}$
agrees to first order in $v/c$ with the last term in  Vekstein's force (\ref{F_V}), disputed in \cite{KMY}, but the term's `dual',
$-(1/c)\,{\bi m}_0\btimes ({\bi v}\bdot \bnabla){\bi E}$, absent in both (\ref{F_V}) and (\ref{F_KMY}), is present, too; the last two terms, where the dots denote time derivatives, are due to possible time dependence of the dipole moments ${\bi m}_0$ and ${\bi d}_0$. Note also that
$\bnabla({\bi d}\bdot{\bi E})=({\bi d}\bdot\bnabla){\bi E}$ when the field ${\bi E}$ is static ($\bnabla\btimes{\bi E}=0$).

The moving particle  acquires  moments  (\ref{d,m}), and thus its interaction with a static field ${\bi E},{\bi B}$ is
\begin{eqnarray}
U&=-{\bi d}\bdot {\bi E}-{\bi m}\bdot {\bi B}\nonumber \\
\fl&=-{\bi d}_0\bdot{\bi E}-\frac{1}{c}\, ({\bi v} \btimes {\bi m}_0)\bdot {\bi E}-{\bi m}_0\bdot{\bi B}+\frac{1}{c}\, ({\bi v} \btimes {\bi d}_0)\bdot {\bi B}.
\label{U}
\end{eqnarray}
The nonrelativistic Lagrangian of such a particle is then
\begin{eqnarray}
\fl L&= \frac{1}{2}\,m v^2 -U\nonumber \\
\fl&=\frac{1}{2}\,m v^2+({\bi d}_0+{\bi v}\btimes {\bi m}_0/c)\bdot {\bi E}+({\bi m}_0-{\bi v} \btimes {\bi d}_0/c)\bdot {\bi B}\nonumber \\
\fl &=\frac{1}{2}\,m v^2+{\bi d}_0\bdot{\bi E}+{\bi m}_0\bdot{\bi B}
+\frac{1}{c}\,{\bi v}\bdot({\bi m}_0\btimes{\bi E}-{\bi d}_0\btimes{\bi B}),
\label{L}
\end{eqnarray}
yielding a canonical momentum
\begin{equation}
\frac{\partial L}{\partial{\bi v}}= m{\bi v}+\frac{1}{c}\,({\bi m}_0\btimes{\bi E}-{\bi d}_0\btimes{\bi B}).
\end{equation}
The use of the Lagrange equation
\begin{equation}
\frac{{\rm d}}{{\rm d}t}\,\frac{\partial L}{\partial{\bi v}} = \frac{\partial L}{\partial{\bi r}}
\end{equation}
with the time derivative
\begin{eqnarray}
\frac{\rm d}{{\rm d}t}\,({\bi m}_0\btimes {\bi E}-{\bi d}_0\btimes {\bi B})
&={\bi m}_0 \btimes ({\bi v}\bdot\bnabla){\bi E}
+\dot{\bi m}_0\btimes {\bi E}\nonumber\\
&\quad-{\bi d}_0 \btimes ({\bi v}\bdot\bnabla){\bi B}
-\dot{\bi d}_0\btimes {\bi B}
\end{eqnarray}
results in a force
\begin{eqnarray}
\fl m\dot{\bi v}
&= \bnabla[({\bi d}_0+{\bi v}\btimes{\bi m}_0/c)\bdot{\bi E}]
+\bnabla[({\bi m}_0-{\bi v}\btimes{\bi d}_0/c)\bdot{\bi B}] \nonumber\\
\fl &\quad
-\frac{1}{c}\,{\bi m}_0\btimes({\bi v}\bdot\bnabla){\bi E}
+\frac{1}{c}\,{\bi d}_0\btimes({\bi v}\bdot\bnabla){\bi B}
-\frac{1}{c}\,\dot{\bi m}_0\btimes {\bi E}+\frac{1}{c}\,\dot{\bi d}_0\btimes {\bi B},
\label{FF}
\end{eqnarray}
which, recalling transformations (\ref{d,m}), is  force (\ref{F}),
understood as mass times acceleration.

Let us now apply expression (\ref{F}) to simplifying examples.
Consider first a constant electric dipole ${\bi d}_0$ with vanishing magnetic moment ${\bi m}_0$, moving with velocity $\bi v$ in a static magnetic field $\bi B$ and zero electric field $\bi E$.  Modelling the dipole as
charges $q$ and $-q$ separated by a small displacement $\bi l$, so that
${\bi d}_0=q{\bi l}$,  the force on the dipole is the net Lorentz force on these charges, which, since ${\bi B}({\bi r}{+}{\bi l}){-}{\bi B}({\bi r})\approx ({\bi l}{\bdot}
\bnabla){\bi B}({\bi r})$, is given by
\begin{equation}
{\bi F}_{\rm L}=\frac{1}{c}\,{\bi v}\btimes ({\bi d}_0\bdot\bnabla){\bi B}.
\label{F_L}
\end{equation}
But according to expression (\ref{F}) with ${\bi E}=0$ and $\dot{\bi d}_0=0$,
the force on the dipole is given as
\begin{eqnarray}
{\bi F}&= \bnabla({\bi m}\bdot{\bi B})
+\frac{1}{c}\,{\bi d}_0\btimes({\bi v}\bdot\bnabla)\bi B\nonumber\\
&=-\frac{1}{c}\,\bnabla[({\bi v}\btimes{\bi d}_0)\bdot{\bi B}]+
\frac{1}{c}\,{\bi d}_0\btimes({\bi v}\bdot\bnabla)\bi B,
\label{F_d}
\end{eqnarray}
where we used ${\bi m}=-{\bi v}\btimes{\bi d}_0/c$.
Let us check whether that agrees with force (\ref{F_L}).
Using standard vector calculus identities, we have
\begin{eqnarray}
\fl-\bnabla[({\bi v}\btimes{\bi d}_0)\bdot
{\bi B}]
&=\bnabla[{\bi d}_0\bdot({\bi v}\btimes
{\bi B})]\nonumber\\
\fl&=({\bi d}_0\bdot\bnabla)({\bi v}\btimes{\bi B})
+{\bi d}_0\btimes[\bnabla\btimes({\bi v}\btimes{\bi B})]
\nonumber\\
\fl&={\bi v}\btimes({\bi d}_0\bdot\bnabla){\bi B}
-{\bi d}_0\btimes ({\bi v}\bdot\bnabla){\bi B}.
\label{F_dd}
\end{eqnarray}
Substituting (\ref{F_dd}) in (\ref{F_d}) yields the Lorentz force (\ref{F_L}).
The term $(1/c){\bi d}_0\btimes({\bi v}\bdot\bnabla)\bi B$, called in question in \cite{KMY}, is thus needed in expression (\ref{F}) to yield the correct force (\ref{F_L}).

Consider further a constant magnetic dipole ${\bi m}_0$ with vanishing electric moment ${\bi d}_0$, moving with velocity $\bi v$ in a static electric field $\bi E$ and zero magnetic field $\bi B$. According to  expression (\ref{F}) with  ${\bi B}=0$
and $\dot{\bi m}_0=0$,  the force on the dipole is now given by
\begin{eqnarray}
{\bi F}&= \bnabla({\bi d}\bdot{\bi E})
-\frac{1}{c}\,{\bi m}_0\btimes({\bi v}\bdot\bnabla)\bi E\nonumber\\
&=\frac{1}{c}\,\bnabla[({\bi v}\btimes{\bi m}_0)\bdot{\bi E}]-
\frac{1}{c}\,{\bi m}_0\btimes({\bi v}\bdot\bnabla)\bi E,
\label{F_m}
\end{eqnarray}
where we used ${\bi d}={\bi v}\btimes{\bi m}_0/c$. Using standard identities,
we have
\begin{eqnarray}
\fl\bnabla[({\bi v}\btimes{\bi m}_0)\bdot{\bi E}]&=
-\bnabla[{\bi m}_0\bdot({\bi v}\btimes{\bi E})]\nonumber\\
\fl&=-({\bi m}_0\bdot\bnabla)({\bi v}\btimes{\bi E})-
{\bi m}_0\btimes[\bnabla\btimes({\bi v}\btimes{\bi E})]\nonumber\\
\fl&=-({\bi m}_0\bdot\bnabla)({\bi v}\btimes{\bi E})
-{\bi m}_0\btimes[(\bnabla\bdot{\bi E}){\bi v}- ({\bi v}\bdot\bnabla){\bi E}].
\label{F_mm}
\end{eqnarray}
Substituting (\ref{F_mm}) with $\bnabla\bdot{\bi E}=4\pi\rho=0$
(the dipole is assumed to be moving in a region where the external charge density $\rho$ vanishes) in (\ref{F_m}) yields
\begin{equation}
{\bi F} =-\frac{1}{c}\,({\bi m}_0\bdot\bnabla)({\bi v}\btimes{\bi E}).
\label{F_AC}
\end{equation}
This is the force on a magnetic dipole moving in a static electric field that is used
in an analysis of Aharonov, Pearle and Vaidman
\cite{APV} of the Aharonov-Casher effect \cite{AC}, which is the electric analogue of the well-known Aharonov-Bohm effect. In that analysis, the force is obtained by invoking the so-called hidden momentum of a magnetic dipole (see e.g.\ \cite{hm} and references therein), but we obtained it here directly using the transformation ${\bi d}={\bi v}\btimes{\bi m}_0/c$ and the Lagrange formalism (a similar procedure has been used in \cite{Al-J}). Without the term $-(1/c){\bi m}_0\btimes({\bi v}\bdot\bnabla)\bi E$ in expression (\ref{F}), the generally accepted quantum-mechanical nature of the Aharonov-Casher effect could be questioned \cite{Boyer}.

Interestingly, Schwinger scattering \cite{Schwing}, which is the scattering of neutrons by the electric field of an atomic nucleus, can be shown to be due to the force (\ref{F_AC}). The Schwinger-scattering Hamiltonian,
\begin{equation}
H=-\frac{\hbar^2}{2m}\,\nabla^2+\frac{{\rm i}\mu\hbar}{mc}\,\mbox{\boldmath$\sigma$}\bdot({\bi E}\btimes\bnabla)
\label{H_Sch}
\end{equation}
(see \cite{LLQED}, equation (42.1)),
is transcribed into a classical Hamiltonian by the replacements $-{\rm i}\hbar\bnabla\rightarrow \bi P$ and
$\mu\mbox{\boldmath$\sigma$}\rightarrow\bi m_0$, where $\bi P$ and $\bi m_0$
are the classical canonical momentum and magnetic dipole moment, respectively:
\begin{eqnarray}
H &=\frac{1}{2m}\,P^2 - \frac{1}{mc}\,{\bi m}_0\bdot({\bi E}\btimes {\bi P})\nonumber\\
&=\frac{1}{2m}\,P^2 - \frac{1}{mc}\,{\bi P}\bdot({\bi m}_0\btimes {\bi E}).
\label{H}
\end{eqnarray}
If this Hamiltonian can be derived from the Lagrangian
\begin{equation}
L= \frac{1}{2}\, m v^2 + \frac{1}{c}\,{\bi v}\bdot({\bi m}_0\btimes {\bi E}),
\label{L_Sch}
\end{equation}
which is that of equation (\ref{L}) with ${\bi d}_0=0$ and ${\bi B}=0$ and thus it yields force (\ref{F_AC}) when $\dot{\bi m}_0=0$, then the force implied by the Hamiltonian (\ref{H}) is the same,
at least to order $v/c$.
But Lagrangian (\ref{L_Sch}) yields a canonical momentum
\begin{equation}
{\bi P}= \frac{\partial L}{\partial\bi v}
= m{\bi v}+\frac{1}{c}\,({\bi m}_0\btimes {\bi E}),
\end{equation}
and thus the Lagrange transformation \cite{LLM} results in a Hamiltonian
\begin{eqnarray}
H= {\bi P}\bdot{\bi v} - L
&=\frac{1}{2}\,mv^2\nonumber\\
&=\frac{1}{2m}\,({\bi P}- {\bi m}_0\btimes {\bi E}/c)^2\nonumber\\
&\approx\frac{1}{2m}\,P^2 - \frac{1}{mc}\,{\bi P}\bdot({\bi m}_0\btimes {\bi E}),
\label{HH}
\end{eqnarray}
dropping in the last line a term proportional to $1/c^2$ in accordance with the fact that the Schwinger Hamiltonian (\ref{H_Sch}) is nonrelativistic. The last line of (\ref{HH}) is indeed the Hamiltonian ({\ref{H}).

\section*{Acknowledgments}
The author acknowledges correspondence with Alexander  Kholmetskii and Oleg Missevitch,
and with Grigory Vekstein, who communicated to the author a preprint of his comment on
\cite{KMY}.
This comment was written by the author in his private capacity. No official support or endorsement by the Centers for Disease Control and Prevention is intended or should be inferred.

\end{document}